\documentclass[a4paper,11pt]{article}
\usepackage{jinstpub} 
\usepackage{lineno}
\nolinenumbers

\title{\boldmath Reporting on pTP sublimation during evaporation deposition}

\author[1]{G. Gomes\note{Corresponding author.},}
\author{B. Gelli,}
\author{V. C. Palavéri,}
\author{R. Sola,}
\author{F. C. Marques,}
\author{E. Kemp}
\affiliation{Instituto de Física \textit{Gleb Wataghin}, Universidade Estadual de Campinas,\\Av. Sérgio Buarque de Holanda 777, 13083-859, Brazil}

\emailAdd{ggomes@ifi.unicamp.br}

\abstract{Noble liquid detectors rely on wavelength shifter materials, such as p-terphenyl (pTP) and Tetraphenyl-butadiene (TPB), which are widely used in neutrino and dark matter experiments. Given their importance, a thorough understanding and characterization of these compounds are essential for optimizing experimental techniques and enhancing detector performance. In this study, we report a novel phenomenon in which commonly used wavelength shifters undergo spontaneous sublimation under high vacuum conditions. We quantify the sublimation rates of pTP and TPB as a function of pressure and temperature, assessing their impact on material growth and physical properties. Additionally, we investigate how variations in film thickness and growth rate influence the sublimation process. These findings provide critical insights into improving the handling and preparation of wavelength shifters during the fabrication of light detectors for these experiments, ensuring their stability and performance in low-background photodetection systems.}


\begin{document}
\maketitle
\flushbottom

\section{Introduction}
\label{sec:intro}

Wavelength shifters (WLS) serve as critical components in noble liquid detectors, since they enable the effective detection of scintillation light by downshifting its wavelength to the visible. Notably, p-Terphenyl (pTP)\footnote{1,4-Diphenylbenzene} and Tetraphenyl-butadiene (TPB)\footnote{1,1,4,4-Tetraphenyl-1,3-butadiene} are among the most widely used WLS materials, and both of them have been the subject of extensive studies about their optical physical properties and relevance in high-energy physics experiments.


The pTP has been extensively utilized in detectors necessitating rapid and efficient wavelength conversion, particularly in combination with liquid argon (LAr). As an example, pTP constitutes an essential component of the far detector in the Deep Underground Neutrino Experiment (DUNE)\cite{abud2024dune}, where it fulfills a significantly important function in converting the 128 nm scintillation light of liquid argon to a longer wavelength, which can be efficiently detected by photo-sensors. The choice of pTP is justified by its high quantum efficiency, brief fluorescence lifetime, and negligible self-absorption, making it ideal for large-scale neutrino detection\cite{acciarri2019design, neumeier2015pterphenyl}.


TPB is commonly used as a wavelength shifter in noble liquid detectors because of its high photoluminescence yield and and its emission is peaked around 420 nm, where traditional photo-detectors are most effective. It has been extensively applied in liquid argon experiments, particularly in the Short-Baseline Neutrino (SBN) program~\cite{machado2019short}, playing a vital role in converting VUV scintillation photons into visible light. TPB's high fluorescence efficiency under cryogenic conditions makes it the preferred choice for experiments needing stable optical performance in harsh environments.~\cite{gehman2011fluorescence, francini2013vuv}.


The selection of appropriate wavelength shifters is key to improving the light-collecting efficiency and, consequently, the overall sensitivity of liquefied noble gases detectors. Ongoing research aims to further improve the stability, efficiency, and durability of these materials under operational conditions, ensuring their continued effectiveness in next-generation neutrino and dark matter experiments.



Understanding the properties these compounds is essential for optimizing experimental methods to enhance light detection. Our research reveals that common wavelength shifters can sublimate in high-vacuum conditions. This study details the sublimation behavior of pTP, quantifying rates at two pressure levels and identifying temperature dependence. We also assessed the impact of sublimation on the materials' physical properties. If this phenomenon is mitigated, material reliability can be improved, lowering experiment costs with sublimated material.

\section{Experimental setup}

\begin{figure}[htbp]
\centering
\includegraphics[width=.4\textwidth]{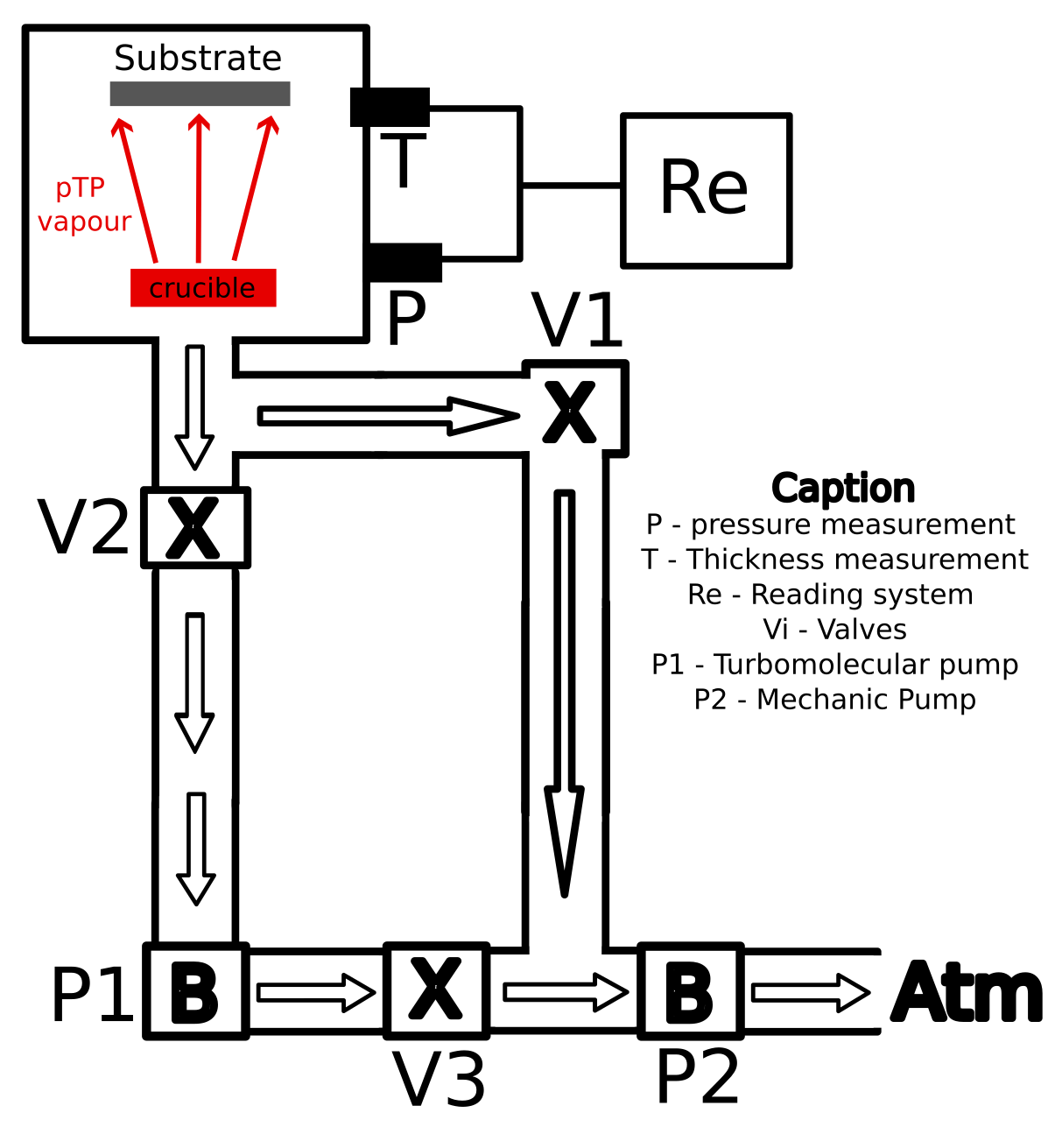}
\caption{Schematic of the thermal evaporation system used.}
    \label{fig:schematic}
\end{figure}

For the development of this work, we used an experimental setup schematically illustrated in Figure~\ref{fig:schematic}. The system consists of a thermal evaporation apparatus capable of reaching pressures as low as $10^{-6}$~Torr. The vacuum system consists of mechanical and turbomolecular pumps connected to the chamber via a valve. The mechanical pump creates the primary vacuum, followed by the turbomolecular pump, which lowers the pressure to high vacuum levels. Inside the chamber, a crucible holds the deposition material. Once high vacuum is reached—critical for sample quality—a 15–20 A current heats the crucible, evaporating the material. The vapor then condenses onto the cooler substrate, forming a uniform layer.

The vapor also accumulates on a quartz sensor, enabling in-situ measurement of the sample thickness. A dedicated monitoring system, designed and implemented by our research group, records the sample thickness at a frequency of 1Hz, along with the chamber pressure. These readings are uploaded to a server, providing high time-resolution data on sample thickness and allowing correlation of pressure variations over time. For this study, we prepared three different samples with varying initial thicknesses and vacuum chamber pressures. The specific characteristics of each sample are presented in Table \ref{tab:samples}. Their thicknesses were monitored over time to evaluate sublimation behavior and support subsequent analysis.

\begin{table}[h]
\centering
\caption{Parameters of the analyzed samples.}
\label{tab:samples}
\begin{tabular}{|c|c|c|c|}
\hline
                                & \textbf{Sample 1}          & \textbf{Sample 2}          & \textbf{Sample 3}          \\ \hline
\textbf{Initial thickness (nm)} & 150                        & 100                        & 50                         \\ \hline
\textbf{Pressure (Torr)}        & 10$^{-6}$ & 10$^{-6}$ & 10$^{-1}$ \\ \hline
\end{tabular}
\end{table}

\section{Results}

\begin{figure}[htbp]
\centering
\includegraphics[width=.4\textwidth]{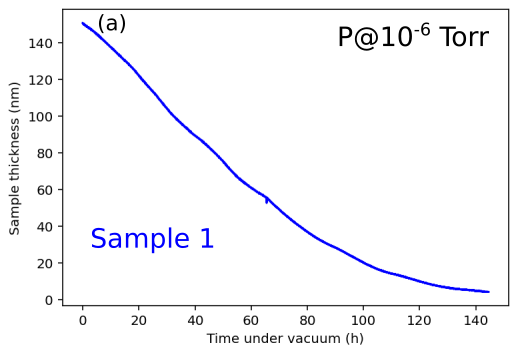}
\qquad
\includegraphics[width=.4\textwidth]{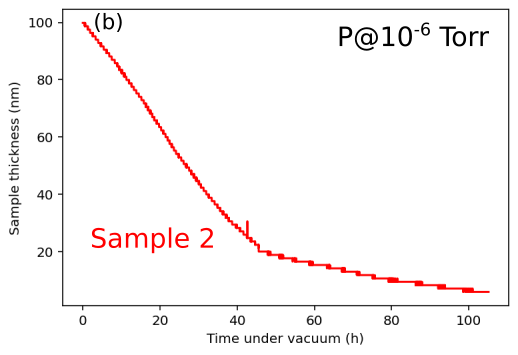}
\qquad
\includegraphics[width=.4\textwidth]{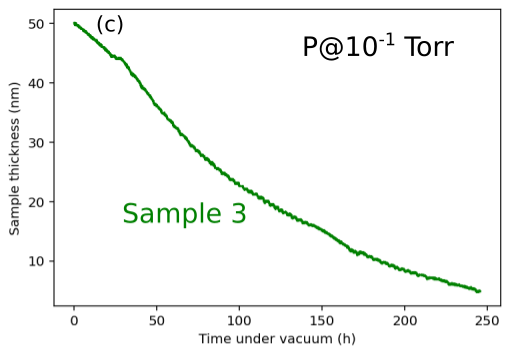}
\caption{In a) the results for Sample 1 and in b) the results for Sample 2 are
shown, both left at a pressure of $10^{-6}$ Torr. In c) Sample 3 is shown, left at
a pressure of $10^{-1}$ Torr}
    \label{fig:thicknesses}
\end{figure}

We analyzed the sublimation rates of three samples. Figure \ref{fig:thicknesses} (a) shows the decreasing thickness of Sample 1 (blue) over time under high vacuum, indicating sublimation. The rate changes when thickness reaches $\sim$~20 nm. This effect may be attributed to interactions between pTP and the substrate.  The average rate for Sample 1 is $-1.50\pm0.22$ nm/h.  Figure \ref{fig:thicknesses} (b) illustrates a similar trend for Sample 2 (red), with a steady thickness decrease and a rate change at 20 nm. Sample 2 sublimates slightly slower than Sample 1, but the difference is within measurement uncertainty. Its average rate is $-1.42\pm0.69$ nm/h, 0.08 nm/h lower than Sample 1. The overall average sublimation rate at $10^{-6}$ Torr is $-1.46\pm0.36$ nm/h. For Sample 3, with an initial 50 nm thickness in a $10^{-1}$ Torr vacuum, Figure \ref{fig:thicknesses} (c) shows a slower sublimation than Sample 1, despite its lower starting thickness. The average rate for Sample 3 is $-0.28\pm0.06$ nm/h, indicative of this pressure.

\begin{figure}[htbp]
\centering
\includegraphics[width=.4\textwidth]{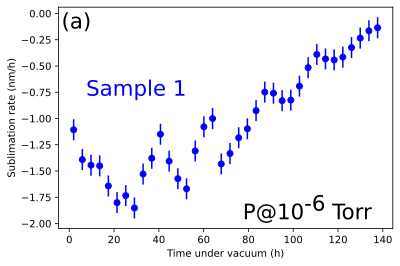}
\qquad
\includegraphics[width=.4\textwidth]{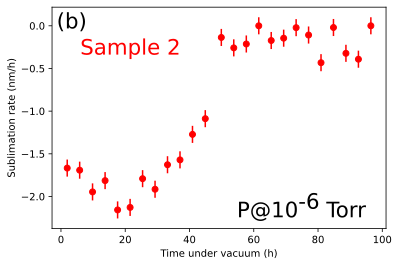}
\qquad
\includegraphics[width=.4\textwidth]{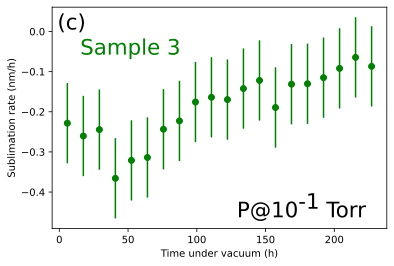}
\caption{In a) the evaporation rates for Sample 1 and in b) the evaporation rates for Sample 2 are shown, both left at a pressure of $10^{-6}$ Torr. In c) Sample 3 is shown, left at
a pressure of $10^{-1}$ Torr}
    \label{fig:rates}
\end{figure}

After determining the sublimation rates for the samples, we performed linear fits on 4-hour segments of Figures~\ref{fig:thicknesses} (a) and (b), and 11-hour segments of Figure~\ref{fig:thicknesses} (c) to analyze how the sublimation rate evolves over time, as shown in Figure~\ref{fig:rates}. For Samples 1 and 2, Figures~\ref{fig:rates} (a) and (b) indicate that the sublimation rate decreases as the sample thickness reaches smaller values, consistent with previous observations. A similar trend is observed for Sample 3 (Figure~\ref{fig:rates} (c)), which was not evident in Figure~\ref{fig:thicknesses} (c). Additionally, Figures~\ref{fig:rates} (b) and (c) point to correlation between thickness and sublimation rate at small thicknesses: the sublimation rate appears to decrease with the thickness of the film, reinforcing our hypothesis that the substrate influences this reduction. However, as shown in Figures~\ref{fig:rates} (a) and (b), the initial sublimation rate for Sample 2 is higher than that of Sample 1, suggesting that additional factors affecting sublimation remain unidentified in this study.


Another characteristic that can be inferred from Figure~\ref{fig:rates} is the temperature dependence of the sublimation process. This figure reveals an oscillation in sublimation rates with a period of approximately 24 hours, which can likely be attributed to daily temperature variations, as temperature was not explicitly measured in this study. Consequently, this provides an initial indication that sublimation is temperature-dependent, as expected. However, this dependence is not verified in the present work and is left for future studies to fully characterize its effects and underlying mechanisms.

\begin{figure}[htbp]
\centering
\includegraphics[width=.8\textwidth]{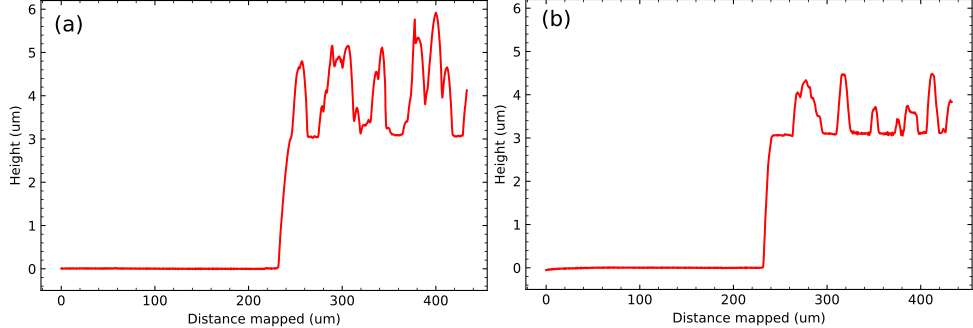}
\caption{Profilometry of a sample grown in this work. (a) shows the sample as grown, while (b) shows the same sample after being exposed to high vacuum for a week.}
    \label{fig:profilometry}
\end{figure}

Finally, Figure~\ref{fig:profilometry} presents profilometry images of the same pTP sample with a thickness of 3~$\mu$m. In (a), the measurement was taken immediately after growth, while in (b), the sample was exposed to high vacuum for one week. The sublimation effects are not visible in the image due to the sample’s high initial thickness. However, one can observe that the grains on the surface, which significantly increase roughness, are noticeably smaller after vacuum exposure. Since sublimation occurs on the surface of the sample, it is expected that the grains sublimate faster than the flat surface underlying because of their larger surface area. As seen in the image, this process leads to a reduction in surface roughness. This phenomenon requires further investigation to better understand the underlying mechanisms and confirm its occurrence, as current evidence remains limited.

\section{Conclusion}


In this study, we identified p-Terphenyl as a volatile compound that undergoes sublimation under high vacuum conditions. Three samples of varying thickness were prepared and subjected to different vacuum levels, aiming to deepen our understanding of its volatility behavior. We quantified the sublimation rates at pressures of $10^{-6}$ and $10^{-1}$ Torr, obtaining values of $1.46\pm0.36$~nm/h and $0.28\pm0.05$~nm/h, respectively. Moreover, our findings indicate a dependency of the sublimation process on sample thickness at smaller scales, presumably due to interactions with the silicon substrate. Preliminary data also suggest the sublimation preferentially impacts grains more significantly than it does flat surfaces. Should this be corroborated, the phenomenon may serve as a viable method for producing samples with reduced surface roughness. Further investigation is necessary to explore the potential influence of temperature on the sublimation process and to ascertain whether similar effects occur in other wavelength shifters, as TPB, for exemple.





\bibliographystyle{JHEP}
\bibliography{biblio.bib}






\end{document}